# Streamer evolution arrest governed amplified AC breakdown strength of graphene and CNT colloids

**Purbarun Dhar [*], Ankur Chattopadhyay [#]**

Department of Mechanical Engineering, Indian Institute of Technology Ropar,

Rupnagar–140001, India

**Lakshmi Sirisha Maganti[#] and A R Harikrishnan [#]**

Department of Mechanical Engineering, Indian Institute of Technology Madras,

Chennai–600036, India

[*] *Corresponding author*: E–mail: purbarun@iitrpr.ac.in, pdhar1990@gmail.com

Phone: +91–1881–24–2173

[#] Equal contribution authors

## Abstract

**The present article experimentally explores the concept of large improving the AC dielectric breakdown strength of insulating mineral oils by the addition of trace amounts of graphene or CNTs to form stable dispersions. The nano-oils infused with these nanostructures of high electronic conductance indicate superior AC dielectric behaviour in terms of augmented breakdown (BD) strength compared to the base oils. Experimental observations of two grades of synthesized graphene and CNT nano-oils show that the nanomaterials not only improve the average breakdown voltage but also significantly improve the reliability and survival probabilities of the oils under AC high voltage stressing. Improvement of the tune of ~ 70-80 % in the AC breakdown voltage of the oils has been obtained via the present concept. The present study examines the reliability of such nano-colloids with the help of two parameter Weibull distribution and the oils show greatly augmented electric field bearing capacity at both standard survival probability values of 5 % and 63.3 %. The fundamental mechanism responsible for such observed outcomes is reasoned to be delayed streamer development and reduced streamer growth rates due to effective electron scavenging by the nanostructures from the ionized liquid insulator. A mathematical model based on the principles of electron scavenging is proposed to quantify the amount of electrons scavenged by the nanostructures. The same is then employed to predict the enhanced AC breakdown voltage and the experimental values are found to match well with the model predictions. The present study can have strong implications in efficient, reliable and safer operation of real life AC power systems.**

## 1. Introduction

The diverse applications of nanomaterials have drawn significant attention in the development of modern and innovative technologies. Among the plethora of nanomaterials discovered in the past few decades, graphene (Gr) and carbon nanotubes (CNT) require no elaborate introduction as these nanomaterials have demonstrated superior and widespread potential as smart systems [1-3]. The innate electronic features of graphene and CNTs, such as ballistic electron transport, high free electron density, semi-metallic behavior, zero gap semiconductors, etc. have been employed in MEMS/NEMS [4, 5], nanoscale electronic transistors, diodes and switches [6-8], power systems such as fuel cells and batteries [9, 10] etc. however, research on the smart features of these carbon based nanomaterials has not been restricted to application in the solid state only. Researchers have examined and implemented graphene and CNTs in the form of stable colloidal suspensions for various applications such as DNA analysis, hazardous material scavenging, water purification, drug delivery [11-14] etc. Also, the enormous thermal transport caliber of graphene and CNTs has attracted attention and several reports have discussed the possibilities of enhancing the thermal properties of fluids by dispersion of such nanomaterials to form colloidal coolants [15-17], both experimentally and through multi-scale modeling approach. However, despite the excellent and tunable electronic transport features of such nanomaterials, it is noteworthy that very less research has been reported on tuning the electrical transport properties of fluids by dispersion of these nanomaterials in colloidal form.

In high voltage power systems, drives and transformers, the major safety and longevity factor for the equipment relies upon the dielectric breakdown strength of the cooling oil which also acts as the liquid insulator. Enhancing the dielectric breakdown strength of such oils is of prime importance from engineering perspective as it ensures safe and reliable operation of the device even in events of voltage surges and spikes in the system, which in general may cause dielectric failure of the oil. Safe operation of modern power systems like transformers, large scale super capacitors, resonator coils, high charge density components etc. requires cooling by insulating oils. To enhance the safety, research and development has been done in the past on addition of organic liquids or additives to the oil in order to improve insulation caliber. Usage of such chemical additives or agents has the innate disadvantages of corrosion of components, possible chemical reactions at high temperatures and electric fields within the devices, chances of catching fire during corona discharge at high voltages failure of the liquids, etc. This has led to the search for alternate additives that do not possess such disadvantages and the materials of choice are nanomaterials [18, 19]. Some of the unique features of nanomaterials that overshadow such disadvantages are high stability as colloids, negligible corrosive action, and high dielectric and electronic properties which can efficiently alter the dielectric characteristics of the oil. In this direction some studies exist where enhanced breakdown (BD) strength employing nanoparticles in insulating fluids is achieved. Semiconducting [20-23] and super paramagnetic [24-26] nanoparticles have been somewhat popular among the works in the domain. The problem that persists is that fail to improve the dielectric breakdown strength at low concentrations. In fact the concentration of particle at which appreciable increment in BD voltage is achieved is not feasible in terms of colloidal stability. At such high concentrations the nano-oil would show signs of sedimentation and clogging of electromechanical components within the power system devices.

Another domain which is less tested is the realistic nature of the testing protocol. In reality, a majority of power systems employing such insulating oils work on alternating current/ voltage systems whereas reported works mostly concentrate on DC systems. Since electrophoresis is very prominent in nanoparticles under DC fields, largely augmented values of dielectric breakdown voltage [18] maybe achieved which in real systems might not be feasible due to presence of AC system where electrophoretic drift is nil. Despite being conducting nanostructures, it is in fact possible that presence of graphene or CNT can lead to large scale augmentation in breakdown voltage of liquid insulators by electron scavenging. A theoretical analysis [27] also discusses that augmentation of BD voltage by scavenging of free electrons is possible for nanostructures with infinitely large electrical conductivities. As the relative electrical conductivity of such nanostructures is around 15 orders of magnitude greater than liquid insulators, such a phenomena is possible based on the theoretical report. The present paper experimentally investigates the augmented AC breakdown strength of Gr and CNT based transformer nano–oils for various concentrations of nanomaterials. In addition to the mean augmentation of breakdown voltage, the reliability or survival probability of such dispersions has been explored with the help of two parameter Weibull distribution analysis. Charge scavenging by the nanostructures has been proposed and mathematically modelled to determine the amount of charge trapped by the nanoparticle population. The model is further extended to predict the enhancement in the AC breakdown strength and the results are found to show good match with experimental values. The mathematical analysis establishes the scavenging model is an accurate theory for the physics behind the breakdown voltage increment mechanism.

## 2. Materials and methodologies

The nanomaterials employed in the investigation constitute two samples of synthesized nano–graphene (Gr) and two samples of procured carbon nanotubes (CNT). The synthesis has been done with commercial reagents which are used as obtained. In general, 1 gm of graphite (G) powder (average flake size ~ 50 μm) is initially added to 25 mL conc. (98 %) sulphuric acid and kept undisturbed at 90 $^o$C for 3 hours. Then 2 gm of potassium peroxosulphate anhydrous and 2 gm of phosphorus pentoxide anhydrous are added and the mixture is stirred vigorously for 3 hours at 90 $^o$C. Preoxidized graphite (pG) is obtained after proper centrifugation of the mixture and it is washed thrice with deionized (DI) water and dried in hot air oven overnight to obtain dry powder form. Further oxidation of the pG to graphite oxide (GO) is done by adding 1 gm pG in 50 mL conc. (98 %) sulphuric acid. The reaction mixture is stirred continuously while maintaining ice cold conditions. To the mixture, 3 gm of potassium permanganate is added in minute batches and the mixture is continuously stirred under ice cold conditions for approximately 6–7 hrs. The ensuing process is highly exothermic and an ice cold atmosphere is mandatory for nullifying explosive reaction. An algae green color (due to formation of manganese dioxide) of the reaction mixture ensures that proper oxidation has occurred. Around 500 mL DI water is added to the mixture and left out till the mixture arrives at room temperature. Finally when 10 mL hydrogen peroxide (5%) is added to the solution, the evolution of a bright yellow solution affirms the formation of GO. The solution is allowed to stand overnight for the GO to sediment. The final precipitate is received after decantation and washed thoroughly with DI water (Millipore). The resulting mixture is again centrifuged at ~ 12000 rpm to obtain the GO, which is then ultrasonicated (Oscar Ultrasonics, India) to produce stable suspensions. The suspensions are then spread uniformly as thin films onto petri dishes and oven–dried overnight. The dried films are scraped off with scalpels to obtain pure, non–aggregated GO.

Next, the dried GO is mixed with DI water (0.1 wt. %) and further ultrasonification yields stable suspensions. To 250 mL of the solution, either 250 mg of finely powdered sodium borohydride or 25 mL of hydrazine hydrate and 10 mL ammonia solution is added and stirred for 1 hr at 60–70 $^oC$. Either of the mentioned processes can be implemented to obtain rGO/Gr suspension from GO. The suspensions are centrifuged at ~ 12000 rpm to obtain the Gr, which is again washed, centrifuged and finally mixed with pure acetone and dried to obtain non–aggregated Gr nano–flakes. A thorough characterization has been carried out to comprehend the physical parameters of the synthesized Gr. The detailed characterizations for the Gr samples have been illustrated in Fig. 1.The SEM image discloses presence of thin flakes, highly convoluted and mimicking folds on sheets of fabric, which indicates the presence of graphenic structures. The oxidation followed by reduction exfoliates the layers, simultaneously creating folds, creases and wrinkles on the individual flakes. Hence, presence of such nanoscale folds and wrinkles (represented in Fig. 1(a) by arrows) qualitatively confirms transformation of graphitic structures to Gr.

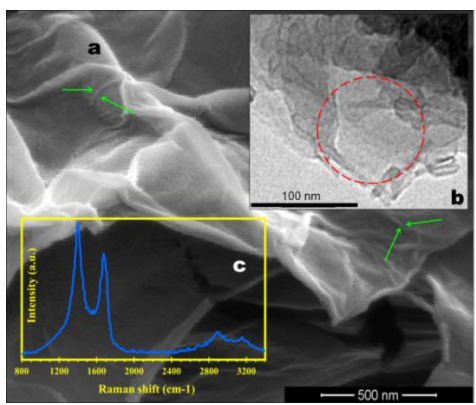

**Figure 1:** (a) HRSEM image of the synthesized graphene samples. The nanoscale wrinkles (shown by arrows) indicate exfoliation of graphite to graphene. (b) TEM image of the graphene sample. Majority of the sample is 2-5 layers thick, thereby allowing for a semi-transparent micrograph (illustrated with circle) (c) The Raman spectra of the graphene samples further indicates graphene with 2-5 layers in general has been formed.

Quantitatively; with the help of Raman spectrum the formation of Gr is ascertained. As illustrated in Fig. 1(c), the spectrum demonstrates the characteristic signature peaks of graphene systems. The D (~1350 $cm^{-1}$) and G (~1600 $cm^{-1}$) bands represent planar defects in the sheets and the in–plane stretching of the $sp^2$ hybridized carbon atoms in Gr respectively. Besides, the existence of the 2D band (~2800 $cm^{-1}$) within the spectrum also endorses the presence of Gr. It is clearly evident that the ratio of intensity of the 2D band to that of the G band for the present samples indicates Gr systems of less than equal to five layers [28], which suggest the unique features of Gr in terms of its augmented performance in the fields of the dielectric and electrical transport. Two distinct Gr samples, viz. Gr1 and Gr2 have been synthesized. The samples are distinguished based on their polydispersity, which has been characterized utilizing DLS (as illustrated in Fig. 1(b)). Analysis reveals that Gr2 has a much larger average flake size than that of Gr1. The CNT systems employed in the present study have been procured from Nanoshel Inc. (USA) (manufactured via Chemical Vapor Deposition (99.5 % purity), illustrated in Fig. 2 and the characteristic dimensions of the nanostructures have been verified utilizing Transmission Electron Microscopy (TEM) and SEM. The TEM images of the CNT samples have been depicted in Fig. 2.

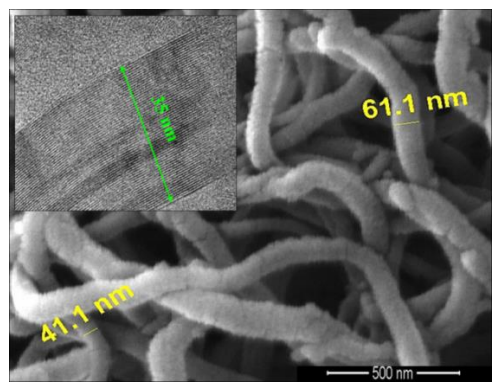

**Figure 2:** HRSEM image of the CNT employed in present article. The inset shows a TEM image of a representative CNT with a relatively few concentric layers. The samples are on and average 10-15 layered.

The nanomaterials have been utilized to prepare nano–oils utilizing two types of industrial transformer oils procured from certified dealers. The nano–oils have been synthesized by adding the required amount of nanomaterial to the oil (in wt. %) followed by ultra-sonication for 1 hour. Oleic acid (OA) has been utilized as the capping agent/ surfactant for samples with nanomaterial concentrations of and above 0.05 wt. % so as to induce more stability to the suspensions. The amount of OA utilized in each case varies from 0.5-2 mL per 500 mL of oil depending upon the concentration of nanomaterials. It is noteworthy that the presence of OA in the oil in such dilute proportions has negligible effect on the breakdown strength of the base oil; as found from experimental observations. Also, the average shelf life of the dilute samples ranges above 2 weeks when kept undisturbed. Illustrations of the base transformer oil, CNT and Gr based nano–oils have been provided in Fig. 3 (c), (d) and

(e) respectively. An automated 50Hz electrical breakdown testing facility has been used to measure AC breakdown voltages in accordance to IEC 60156 codes employing an acrylic test basin and brass spherical electrodes set at 2.5mm gap. A typical setup of the experiments is illustrated in Fig. 3. The rate of growth of the voltage has been maintained linear and fixed at 2 kV/s as per standardized protocol. An initial standing time of 5 minutes has been allowed before the application of voltage in order to avoid effects of stray bulk motion within the fluid. During each breakdown event, coronary discharge produces black residue between the electrodes. The nano-oil samples are stirred with an inert glass rod after each test to nullify any additional effects of concentrated carbon residue present between the electrodes. The oil sample is changed after 5 breakdown events to eliminate chances of contamination due to the carbonaceous residue formed during discharge. All experiments are performed at room temperature. Four sets of breakdowns have been carried out for all oil samples and in total twenty readings were obtained for each sample.

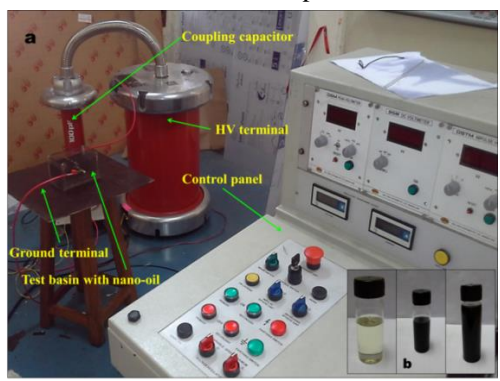

**Figure 3:** (a) Illustration of the experimental test rig (b) Representative transformer oil and nano-oil samples.

## 3. Results and discussions

The presence of nanoparticles in the oil yields augmented dielectric breakdown (BD) performance in terms of breakdown strength, both the average value as well as the statistical reliability value. The BD characteristics of Gr samples (G1 and G2) based oil has been illustrated in Fig. 4. Experimental observations reflect that using stable suspension of Gr nano-oils intensify the capability of the oil to withstand higher voltages before it loses its insulating capacity due to ionization induced streamer formation and merging to create breakdown event. From Fig. 4(a) it can be observed that with increasing the concentration of the nanoparticulate phase, BD strength enhances steadily. The rise is initially linear but slowly begins to attain a plateau. The concentration at which highest value of BD voltage is obtained hereby referred as the critical particle concentration (CPC). Further addition of Gr beyond CPC tends to reduce its BD strength gradually. Therefore, the present study concentrates around region nearby the CPC in order to ensure that the highest possible benefit in terms of BD voltage augmentation is obtained. Since graphene and CNT are structures that tend to form percolation networks in colloidal [15], the CPC is the limit beyond which these percolation networks manifest due to enhanced nanomaterial population. The presence of such highly conducting percolation chains in between the electrodes often acts as the electron conduction pathway. In such cases the breakdown occurs due to electron hopping via the percolation paths rather than streamer merging [18]. Accordingly, beyond a certain point, further addition of such particles deteriorates breakdown strength. An analysis comprising of the relative enhanced in BD strength between the two samples of Gr has also been presented in Fig. 4(b) for exploring nanomaterial size effects. Both the samples follow similar trend in terms of BD strength nature, however, the maximum possible value of augmentation noticed for G1 is slightly higher than that of G2. The examination of the experimental data unveils that both G1 and G2 nano-oils experiences the highest BD value at almost similar CPC value. This thereby further cements the fact that the deterioration in performance beyond this point arises due to percolating structures rather than material or size features. Considering the maximum BD value measured from the experiment, it can be proposed that ~ 30-40% of enhancement of the BD voltage is achievable by allowing trace amounts of such nanomaterials in base oils. However, beyond a concentration limit of 0.1 wt. %, the results indicate addition of Gr does not demonstrate any appreciable improvement, especially in case of G1.

Although Fig. 4 illustrates the enhancement of BD characteristics in terms of the average dielectric strength, it is often more important to portray the capability of an insulator to withstand electrical stress based on a statistical reliability. The complete set of data obtained from repetitive experimentation for a particular sample of oil is used to generate two a parameter Weibull distribution. The two parameters, namely the shape and scale factors, are derived by fitting the data set to a Weibull distribution for the different nano-oils. The parameters are further employed to obtain the survival probability or the reliability of the nano-oils at a given value of electric field intensity. Fig. 5(a) illustrates the reliability curves for G1 based oil samples with respect to electric field intensity. Any point on a curve represents the statistical probability of survival for that particular oil sample at that particular electric field intensity. The G1 based oil samples are selected for since the discussion in the previous paragraph shows that G1 based samples show higher enhancements and hence the highest achievable reliability is sought in the present context. For a given electric field strength and selected particle concentration; the propensity of the nano-oil failure can be estimated from Fig. 5 (a).

Often in reliability assessment using a twin parameter model, two fixed values of the failure probability are employed to further judge the behavior of the sample. The failure probabilities of 5 % and 63.3 % are used in the exercise as standard values. The failure probability values for each nano-oil sample have been illustrated in Fig. 5 (b). The method to interpret data from the figure is discussed briefly. Considering the base oil case from Fig. 5 (b), it signifies that at 4 kV electric field the oil has a failure probability of 5 % whereas when the field value is raised to 5 kV; it has a breakdown failure probability of 63.3 %. It is important to observe from the figure that the 5 % failure probability level enhances with increasing Gr concentration which indicates that the addition of nanomaterials increases the upper bound survival level. It is also noteworthy that the difference between the 5 % failure level and 63.3 % levels also increase as particle concentration is increased. This indicates that even the upper bound of failure is delayed by the addition of nanoparticles and increasing the concentration improves the delay. Hence, while the base oil has a 63.3 % chance to undergo breakdown at ~ 5 kV, addition of 0.05 wt. % Gr ensures that the nano-oil has a 63.3 % failure chance only when the voltage is ~ 11 kV.

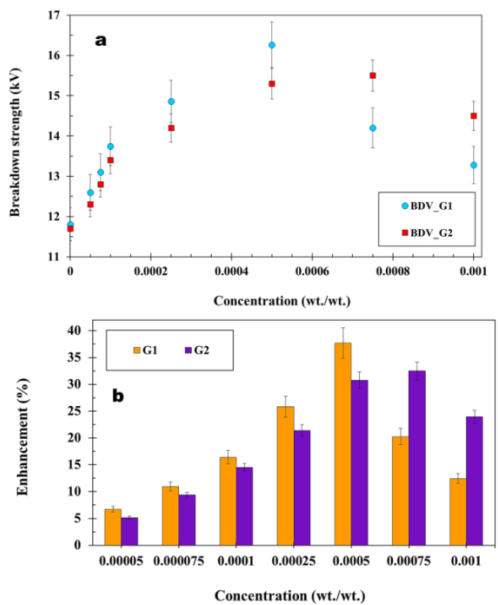

**Figure 4:** (a) AC breakdown strength with respect to graphene concentration and sample (b) The corresponding enhancement in the breakdown strength with respect to the base oil.

The mechanism of enhanced dielectric BD strength due to addition of nanomaterials can be explained based on the electrohydrodynamics of the colloidal phase under electrical stressing [18]. A theoretical model [27] describes the possible phenomenon responsible behind augmentation of BD voltage based on electron scavenging by the nanomaterials of high electrical conductivity. Under electric stress, the oil molecules near the electrodes ionize to form charged entities. These charge carriers accumulate near the electrodes to form two streamers that grow with increasing electric field. Once the streamers are saturated with charged units and grow in size to merge, a conduction pathway is created from one electrode to the other through the ionized oil. This leads to a corona discharge across the electrode indicating dielectric breakdown. The uniformly distributed nanostructures reduce the streamer propagation velocity and growth rate by trapping the rapidly travelling electrons formed due to field ionization. This process is called scavenging and consequently the growth of streamers is retarded, resulting in higher BD strength. The size effects of nanoparticles in oils also reveal that smaller size particles have larger surface area to volume ratio and thus have the capability of arresting more number of free electrons per unit volume [29], thereby resulting in decelerated growth of streamers and further enhancing the BD strength.

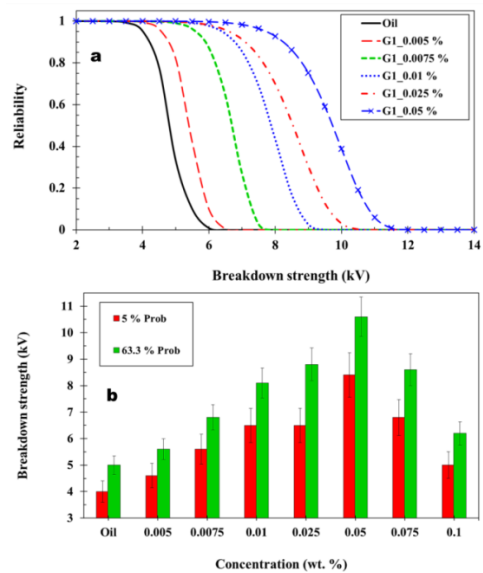

**Figure 5:** (a) Reliability plot with respect to breakdown voltage for G1 based oils as derived from twin parameter Weibull analysis (b) The corresponding breakdown voltages as function of graphene concentration for typical 5 % and 63.3 % failure probabilities.

In fact the presence of nanoparticles leading to effective scavenging of free electrons can also be testified via observation the event of corona discharge. Fig. 6 illustrates the images of corona discharge during breakdown event for different concentrations of the Gr nanoparticles in oil. Corona is characterized by high current high voltage electric jolt which appears when excessive the conduction pathway opens up due to complete ionization of the oil. It is accompanied by emission of light and sound energies due to its intense energy. It is evident from Fig. 6 that the energy released in terms of light during breakdown diminished with addition of nanoparticles and the audible crack during the discharge event also dims down to certain extent. Essentially, in spite of the nano-oils undergoing failure at higher voltages, the discharge phenomena is less violent. These observations clearly reflect the diminishing nature of streamer growth which plays dominant role in the BD event. It can be inferred from Fig. 6 that increased amount of nanoparticles delay the streamer growth process by electron scavenging, leading to a reduced population of electrons available to lead to the BD event. Since the total available charge to initiate breakdown is diminished, the corona discharge occurs at lower energies, thereby releasing less light and sound energies. This is also important from an engineering perspective as it leads to improved life of the electrical parts in actual devices which would otherwise be damaged after few breakdown events due to the massive energy released during the corona discharge.

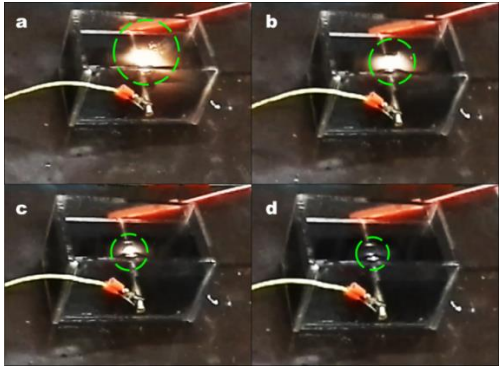

**Figure 6:** Reduction in corona discharge/ arc flash intensity and strength as function of increasing graphene concentration indicating electron scavenging by the nanostructures (a) base oil (b) 0.01 wt. % (c) 0.025 wt. % and (d) 0.05 wt. %.

It can be observed that use of CNT as a nanomaterial in base oils influences its performance by developing higher resistance to dielectric breakdown and in several cases the enhancement is higher compared to Gr based oils of same concentration. Fig. 7(a) illustrates that inclusion of CNT decelerates the breakdown mechanism by retarding the streamer velocity of nano-oil, leading to augmented dielectric BD voltage. Coherence in augmentation between the two samples of CNTs employed (CNT1 and CNT2) can be ascertained from Fig. 7(b) where the percentage improvement in augmentation has been shown. It is noteworthy to mention that both of these samples generate the maximum enhancement for the same value of CPC. In fact the receding trend beyond CPC also confirms similar behavior between the samples which further establishes that the CPC is a structural percolation based phenomena rather than a material property. However, despite the larger aspect ratios of CNTs compared to Gr, the CPC is obtained at higher concentrations in case of CNTs. Given the fact that the electronic conductivity of Gr is slightly higher compared to CNTs, the electron hopping mechanism is stronger in Gr than CNT, thereby causing earlier percolation induced BD despite having smaller length scale compared to CPC. Thereby, it may be reasoned that both percolation propensity as well as the electronic conductance are important to determine the CPC for a material in such scenarios. It is observed that the enhancement in BD voltage for CNT1 based oil is close to 70% at its peak; almost double than that of the Gr1 based oil. The reliability of CNT1 based oil, as determined from the Weibull distribution is represented against BD voltage in Fig. 8 (a). It can be observed that 0.05 wt. % yields the highest magnitude of reliability among all of the particle concentrations, both for 5 % as well as 63.3 % failure estimates.

The caliber of a nanomaterial in increasing breakdown strength of a non-polar liquid can be made based on the charge relaxation time ($\tau$), absolute permittivity ($\varepsilon$) and electrical conductivity ($\sigma$) of the nanostructure (subscripted $np$) with respect to the surrounding fluid medium (subscripted $f$) [18]. Initially, a nanostructure (of characteristic length '$d_{np}$') situated at a location at $(r,\theta,z)$ coordinates at time ($t$) is considered, such that the origin of the polar coordinate system coincides with the center of the nanostructure. An electric field ($E_0$), with the nano–oil as the medium, is applied between the working electrodes. In absence of nanostructures, the field is expected to distribute uniformly within the oil but the presence of nanostructures causes local disruption of the uniform field near the nanostructures. The newly distributed electric field may be expressed as a function of the $(r,\theta)$ location as [27]

$$E_r(r,\theta) = E_0 \cos\theta \left[ 1 + \left\{ \left(\frac{d_{np}^3}{4r^3}\right)\left(\frac{\varepsilon_{np}-\varepsilon_f}{\varepsilon_{np}+2\varepsilon_f}\right) e^{-t/\tau} \right\} + \left\{ \left(\frac{d_{np}^3}{4r^3}\right)\left(\frac{\sigma_{np}-\sigma_f}{\sigma_{np}+2\sigma_f}\right)(1-e^{-t/\tau}) \right\} \right]$$
(1)

$$E_\theta(r,\theta) = E_0 \sin\theta \left[ -1 + \left\{ \left(\frac{d_{np}^3}{8r^3}\right)\left(\frac{\varepsilon_{np}-\varepsilon_f}{\varepsilon_{np}+2\varepsilon_f}\right) e^{-t/\tau} \right\} + \left\{ \left(\frac{d_{np}^3}{8r^3}\right)\left(\frac{\sigma_{np}-\sigma_f}{\sigma_{np}+2\sigma_f}\right)(1-e^{-t/\tau}) \right\} \right]$$
(2)

With the charge relaxation time is expressed as

$$\tau = \frac{2\varepsilon_{np}+\varepsilon_f}{2\sigma_{np}+\sigma_f} \quad (3)$$

Considering the graphene and CNT based oils, the relaxation times may be estimated. The relative permittivity of the oil is found using using a multi–frequency LCR meter (HP, USA). At frequency range of 10–100 Hz (frequencies chosen as per distribution parameters worldwide, i.e. 50 or 60 Hz), the oil has a relative permittivity of ~2.2. The electrical conductivity of the oil is ~ $10^{-12}$ S/m (measured using an electrical conductivity probe). The relative permittivity of Graphene and CNT also tend towards metallic systems, with values ranging from 2–2.5 in general but with electronic conductivities of the order of $10^8$ S/m. Based on the data, the magnitude of '$\tau$' for the nano-oils may be predicted to be ~ $10^{-19}$ s. However, depending on composition and dielectric properties, the characteristic time scale for streamer growth in mineral oils ranges from $10^{-10}$ s to $10^{-6}$ s [27]. Hence, the charge relaxation time for the nanomaterials in the oil is at the very least 10 orders of magnitude smaller than the streamer development time scale. Thus the nanomaterials are able to scavenge the released electrons much faster than the time required by the electrons to add to the streamer growth process. Consequently, the probability of DB at the expected value is drastically reduced, and the oil naturally achieves higher electrical stress bearing caliber.

Since graphene and CNT have large electrical conductivities, Eqn. 1 can be restructured for infinitely conducting nanomaterials assumption to yield a reduced form as [27]

$$E_r(r = d_{np}/2, \theta) = 3E_0 \cos\theta \quad (4)$$

As evident, the maximum value of the electric field localized to the nanostructures is 3 times compared to the field existing across the electrodes. Hence the ionized molecules in and around the local neighborhood of such nanostructures experience a higher electrostatic force towards them compared to that of the electrodes. Accordingly, a large population of charged entities is unable to develop into the streamer and is rather scavenged by the nanomaterials at higher electric potential. The enhanced electric field in the vicinity of the nanostructures is also a potential reason for highly enhanced breakdown strength even for dilute nanostructure based oils. As the electric field intensity across the electrodes increase, more and more oil molecules near the electrodes are stripped down into electrons and ionized atoms by the electric stress. These charges units form the streamers around the electrodes, which eventually merge, creating a conduction pathway and leads to breakdown. The graphene and CNT nanostructures scavenge these charge carriers, thereby delaying streamer growth and breakdown. However, each individual nanostructure has a finite ability to scavenge charges before reaching its saturation scavenging limit. As the nanostructures scavenge charged units; the effective charge on the nanostructures increases. This leads to repulsion of the incoming charges, thereby reducing the scavenging efficiency with time and finally leading to a saturation capacity for scavenging. The net effective charge $Q(t)$ on the nanostructure increases with time and thereby modifies the electric field in its locality. The radial and polar components of this localized field near a nanostructure can be expressed as

$$E_r = E_0 \left[ \left(1 + \frac{d_{np}^3}{4r^3}\right) \cos\theta + \frac{Q(t)}{\pi \varepsilon_f d_{np}^2} \right] \quad (5)$$

$$E_\theta = E_0 \left(-1 + \frac{d_{np}^3}{8r^3}\right) \sin\theta \quad (6)$$

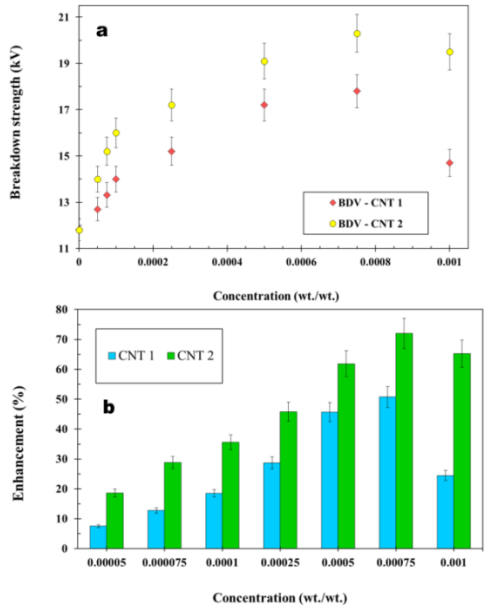

**Figure 7:** (a) AC breakdown strength with respect to CNT concentration and sample (b) The corresponding enhancement in the breakdown strength with respect to the base oil.

Electron scavenging by the nanostructures is possible as long as the net effective potential at the nanostructure surface is positive. Once the effective potential becomes negative due to scavenging of electrons, the nanostructure can no further attract incoming electrons Speaking mathematically, for scavenging, the potential at distance equal to nanostructure radius from its center (r=$d_{np}$/2) must be finitely positive ($E_r \geq 0$). Inserting the said condition, Eqn. (5) the requisite condition for scavenging becomes

$$\left| \frac{Q(t)}{3\pi \varepsilon_f d_{np}^2 E_0} \right| \leq \cos\theta \quad (7)$$

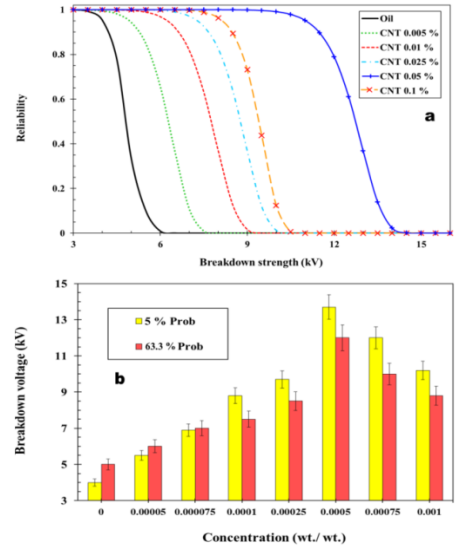

**Figure 8:** (a) Reliability plot with respect to breakdown voltage for CNT1 based oils as derived from two parameter Weibull analysis (b) The corresponding breakdown voltages as function of CNT concentration for typical 5 % and 63.3 % failure probabilities.

Here $\theta$ represents the polar window available for scavenging. At the very moment the potential at nanostructure surface becomes negative, further scavenging terminates. The further ionized electrons then form the streamer and cause breakdown of the oil. Once the potential becomes zero, the polar window available for scavenging also dies out, i.e. mathematically it is characterized by $\theta=0$. Hence, the net charge scavenge by the nanostructure until this point may be denoted as the saturation scavenging caliber ($Q_{S,np}$) for the nanostructure and is expressed as

$$Q_{S,np} = 3\pi \varepsilon_{eff} d_{np}^2 E_0 \quad (8)$$

The expression employs an equivalent particle diameter for the nanostructures used. The methodology to compute the equivalent spherical diameter for a CNT or a graphene flake has been described in published literature [18]. Eqn. 9 also uses the effective dielectric constant for the colloid which may be computed from the Matijevic equation [18] expressed in Eqn. (9) based on the permittivities of the fluid and nanomaterial and nanomaterial concentration ($\varphi$, wt./wt.).

$$\varepsilon_{eff} = \varepsilon_f + \frac{3\varepsilon_f(\varepsilon_{np} - \varepsilon_f)}{2\varepsilon_f + \varepsilon_{np}}\varphi \qquad (9)$$

Eqn. 8 describes the saturation scavenging limit for a single nanostructure. The number density of nanostructures ($n_{np}$) in the medium can be deduced from the nanomaterial concentration, densities ($\rho$) of the fluid and nanomaterial and volume of a single nanostructure ($v_{snp}$) as

$$n_{np} = \frac{\rho_f \varphi}{\rho_{np} v_{snp}} \qquad (10)$$

During ionization, the electrons are pulled off the oil molecules. Accordingly the molecules attain positive charges, which lead to formation of independent positive streamers and may also lead to breakdown [27]. As the nanostructures gain electrons, their net positve potential reduces, thereby attracting them to the positive streamers. The diffusive migration of the nanostructures into the positive streamers also hampers the steady growth of these streamers, further delaying breakdown. Also Brownian fluctuations by the nanostructures disrupt the process of steady streamer formation. Hence the couple electro-thermal diffusive mobility of the nanostructure is largely responsible in delaying positive streamer growth. The electro–thermal mobility ($\mu_{e-t}$) is expressed in terms of the acquired charge, intrinsic thermal energy (product of the Boltzmann constant $k_B$ and absolute temperature $T$) and Brownian diffusivity ($D_B$) as [18]

$$\mu_{e-t} = \frac{Q_{s,np} D_B}{k_B T} \qquad (11)$$

In order to determine the electro thermal diffusivity of the nanostructures within the fluid, the ratio of its maximum charge content and thermal diffusion across the viscous fluid medium needs to be invoked. Due to their sizes, nanostructures can be accurately assumed to follow Stokesian dynamics in a fluid system. Hence the Stokes-Einstein Brownian model for Brownian diffusivity [18] can be used to obtain the expression for the electro–thermal mobility of a unit nanostructure as

$$\mu_{e-t} = \frac{|Q_{s,np}|}{3\pi \eta d_{np}} \qquad (12)$$

In the above equation, the diameter of nanoparticle should be replaced by the equivalent diameter for Gr and CNT as discussed earlier. In reality, the capability of the nanostructure to hamper the positive streamer growth due to diffusive migration is governed by the order by which its own diffusivity is stronger than the positive streamer diffusivity ($\mu_C$, in general ~ $1\times10^{-9}$ m$^2$V$^{-1}$s$^{-1}$ [27]). Hence the ratio of the two diffusivities governs the overall mechanism. The net charge scavenged ($\chi_{np}$) from the streamers by the total involved nanostructure population up to saturation is thus expressed as

$$\chi_{np} = n_{np} Q_{S,np} \left( \frac{\mu_{e-t}}{\mu_c} \right) \qquad (13)$$

Breakdown in insulators occur when the electric stress leads to excessive ionization of the molecules, leading to an avalanche of free charge carriers to stream across the material, momentarily creating a conducting pathway through it. Since molecular ionization is involved, the breakdown voltage may be expected to hold an inversely proportional relationship to the maximum free charge density, i.e. the total free charge per unit volume which can be stripped off from its bound form under electrical stress. The scavenging action by the nanomaterials delays the event by reducing the rate of growth of free charge population. The enhancement in breakdown voltage is thereby directly linked to the quantum of free charges scavenged by the nanoparticle population. The breakdown in case of such nano-oils thus only occurs when the nanomaterials are no further able to scavenge, or have achieved their saturation charge scavenging caliber. The saturation charge population that the can be scavenged by the nanostructures can be determined from Eqn. 12. Based on the maximum free charge density of the base fluid ($\chi_f$) (~ 450 C/m$^3$ for transformer oils [27]), the probability ($p(b)$) by which breakdown event at the expected point has been delayed is expressed as [18]

$$p(b) = \frac{\chi_{np}}{\chi_f} \qquad (14)$$

Eqn. (14) provides an estimate of the probability by which the oil survives the breakdown process at the intended strength due to presence of nanostructures. Thus this probability when infused to the expected breakdown strength would yield the enhanced breakdown strength. Assuming the enhancement may be assumed to behave linearly, the enhanced breakdown voltage ($V_e$) is expressible as

$$V_e = V_f (1 + p(b)) \qquad (15)$$

The predicted values from Eqn. 14 have been compared with respect to experimental observations for graphene and CNT based oils and illustrated in Fig. 9 (a) and (b) respectively. It is seen that the predictions are in general accurate to the critical particle concentration. Beyond the critical point, the higher concentration leads to percolation chains of graphene or CNT being formed. This often bridges the gap between the electrodes and the streamers are able to propagate faster across the percolation network, leading to reduced breakdown strength [18].

## 4. Conclusions

To infer, the present article explores the idea of enhancing the AC dielectric breakdown strength of insulating mineral oils by adding trace amounts of Gr or CNTs to for stable dispersions. The concept of incorporating infinitely conducting nanostructures in insulating liquids to improve the AC breakdown strength is taken from reported mathematical analysis [27]. The relatively huge magnitude of electrical conductivities of Gr and CNTs compared to such mineral oils allows for these nanostructures to be assumed to be infinitely conducting. Detailed experiments have been carried out employing an AC high voltage facility to understand the increment in the BD strength of the oils with respect to nanostructure concentration. It is observed that the AC breakdown strength of such oils can be improved by ~ 50 % and ~ 70 % by infusing with Gr and CNTs respectively, at concentrations as low as 0.05 wt. %. Furthermore, the statistical reliability of the nano-oils has been determined using a two parameter Weibull analysis to best fit the data.

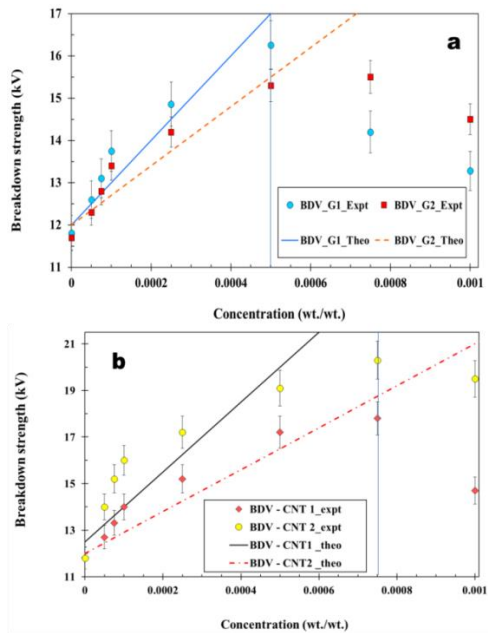

**Figure 9:** Predictions from the proposed theory with respect to the experimental data for (a) graphene and (b) CNT based oils. The predictions are valid only in the region of enhancing breakdown strength and cannot capture the deterioration beyond the critical concentration. The vertical lines illustrate the critical zone.

The shape and spread factor for the data set is obtained from the Weibull distribution fit and the same is further employed to determine the reliability or the survival probability of the oils with respect to increasing electrical stressing. It is observed that addition of such nanostructures not only improves the average AC breakdown strength but also enhances the reliability of the oil over a broad range of electrical field intensities. Scavenging of the electrons from the electrically stressed oils by the nanostructures has been proposed as the possible mechanism for delay of the streamer development and enhancement of BD strength. Based on the same concept, a mathematical model to predict the amount of ionized electrons scavenged by the nanomaterials is put forward in the present paper. Based on the amount of charge scavenged, the increment in the AC breakdown strength is mathematically determined and it is observed to predict the experimental observations with good accuracy. The present approach and findings may have important implications on improving safety and reliability of power equipment by enhancing the AC dielectric strength of the involved liquid coolants and insulators.

## Acknowledgments
The authors thank Dr. C. C. Reddy of the Department of Electrical Engineering of IIT Ropar for access to the high voltage laboratory facilities and Mr. Satyamoorthy of IIT Ropar for guidance and assistance. PD also thanks IIT Ropar for financial support towards the present work (grant number IITRPR/Research/193)